# ContactLess Integrated Photonic Probe for light monitoring in InP-based devices

**Daniele Melati**, **Marco Carminati, Stefano Grillanda, Giorgio Ferrari, Francesco Morichetti, Marco Sampietro and Andrea Melloni**

*Dipartimento di Elettronica, Informazione e Bioingegneria, Politecnico di Milano, via Ponzio 34/5, Milano*

daniele.melati@polimi.it

**Abstract:** The increasing complexity of photonic integrated circuits requires the possibility to monitor the state of the circuit in order to stabilize the working point against environmental fluctuations or to perform reliable reconfiguration procedures. Although InP technologies can naturally integrate high-quality photodiodes, their use as tap monitors necessarily affects the circuit response and is restricted to few units per chip. They are hence unsuited for very large circuits, where transparent power monitors become key components. In this paper we present the implementation of a ContactLess Integrated Photonic Probe (CLIPP) realizing a non invasive integrated light monitor on InP-based devices. We describe an innovative vertical scheme of the CLIPP monitor which exploits the back side of the chip as a common electrode, thus enabling a reduction of the device footprint and a simplification of the electrical connectivity. We characterize the response of the CLIPP and demonstrate its functionality as power monitor. Lastly, we provide a direct demonstration that CLIPP monitor allows to access more accurate information on the working point of photonic integrated circuits compared to conventional external or integrated photodetectors.

## 1. Introduction

Since the early beginning of photonics there has been a continuous and constant development towards very large scale of integration [1]. A pressing request that brought unavoidably to a rapid shrink of the physical dimensions of each integrated device. In particular, the investigation of high index contrast technologies allowed to accommodate a larger number of components on small areas and consequently gave rise to complex and innovative circuits [2,3].

On the other hand, the realization and use of advanced circuits poses immediately the problem of control and stabilization, since fabrication tolerances, fluctuations and crosstalk (thermal, electrical or optical) can continuously change the optimal working point of the circuit [4,5]. Real-time local monitoring is hence needed to counteract these spurious effects and lock the circuit at the desired state [6]. Moreover, the techniques to actively control and change the circuit responses [7] would largely benefit from the possibility to easily and dynamically probe the working point of the circuits.

For Indium Phosphide-based technological platforms the possibility to monolithically integrate complex circuits comprising sources, amplifiers and detectors together with passive circuitry as power splitters, waveguides or filters makes the problem of control a particularly critical aspect. This issue is classically solved by tapping part of the power from waveguides in few selected points and exploiting photodiodes, easily integrated in InP technologies [1]. This approach has an intrinsic limitation in the number of tapping points that can be placed in a single circuit without severely affecting its functionality. Power tapping and photodiodes cannot hence be a solution when the complexity grows to hundreds or thousands of single building blocks, where non-invasive light monitoring becomes an essential feature. Further, transparent monitoring of light in optical waveguides would enable the realization of complex functionalities not achievable with conventional detectors [8].

In previous works [9,10] we recently proposed a ContactLess Integrated Photonic Probe (CLIPP) as an innovative and viable solution to overcome the problem of transparent light monitoring on Silicon on Insulator (SOI) platforms. The CLIPP monitors the light dependent variations of the electric conductivity of the semiconductor waveguide core that are induced by a surface-states carrier generation effects [9]. A capacitive access to the waveguide is exploited avoiding a direct contact between the CLIPP and the waveguide core, ensuring transparent operation.

In this work we demonstrate that the CLIPP can be implemented also in InP-based technologies. We also introduce an innovative geometry for the CLIPP that uses a vertical rather than co-planar approach. The very small (although finite) conductivity of the passive semi-insulating InP substrate allows to contact directly the back side of the chip, used as common a electrode. Since a single metal strip on the waveguide is required instead of an electrode pair [9,10], this geometry would enable a reduction of the overall CLIPP size. Finally, the CLIPP is realized without any modification to the existing technological platform offered by the foundry, thus making it suitable for the introduction in a generic foundry scheme.

The paper is organized as follows. Section 2 describes the realization of the CLIPP device on InP-based waveguides. The vertical sensing approach is presented along with its equivalent electrical model. The characterization of the sensor response and its application as in-line power monitor are presented in section 3. Lastly, section 4 reports on the results obtained in the extraction and analysis of the transfer function of a ring resonator coupled to an external Fabry-Pérot cavity using a CLIPP directly placed within the ring cavity.

## 2. The CLIPP device on InP technology: vertical sensing approach

The InP-based optical waveguide exploited for the fabrication of the device is a 2 µm-wide rib-shaped waveguide, whose scheme is shown in figure 1(a). The InGaAsP core is 1µm-high with an etch depth of 600 nm. The InP substrate has a thickness of more than 250 µm. The entire layer stack is highly Fe-doped, guaranteeing a semi-insulating behavior with a carrier concentration comprised between $10^7$ $cm^{-3}$ to $10^8$ $cm^{-3}$. The small residual conductivity of the stack allows to realize the CLIPP exploiting a vertical rather than coplanar approach. The bottom of the chip, which is used as a large common electrode, is directly contacted through a metallic holder and does not require a dedicated metallization of the back side of the sample. The second metal electrode of the CLIPP seats on the top of the waveguide. A SiN isolation layer is placed between the waveguide core and the metal strip, but any other electrically insulating layer (e.g. BCB polymer) would provide the required capacitive access to the waveguide.

The realized device [figure 1(b,c)] is an all-pass ring resonator with physical length of 3.763 mm. A Two Mode Interference coupler [11] is used to couple the ring to the bus waveguide with a coupling factor comprised between 20% and 25%, as measured in dedicated test structures. The top electrode of the CLIPP is placed inside the ring resonator and has a length L of about 2 mm. Although several pads are present along the metal strip, only the top one was used to contact the electrode by means of a micro probe. With this scheme the CLIPP can directly monitor the optical power circulating in the cavity. As demonstrated in the next sections, this allows to monitor variations of the amount of power coupled in the bus waveguide through the input fiber and also to characterize the optical transfer function of the ring.

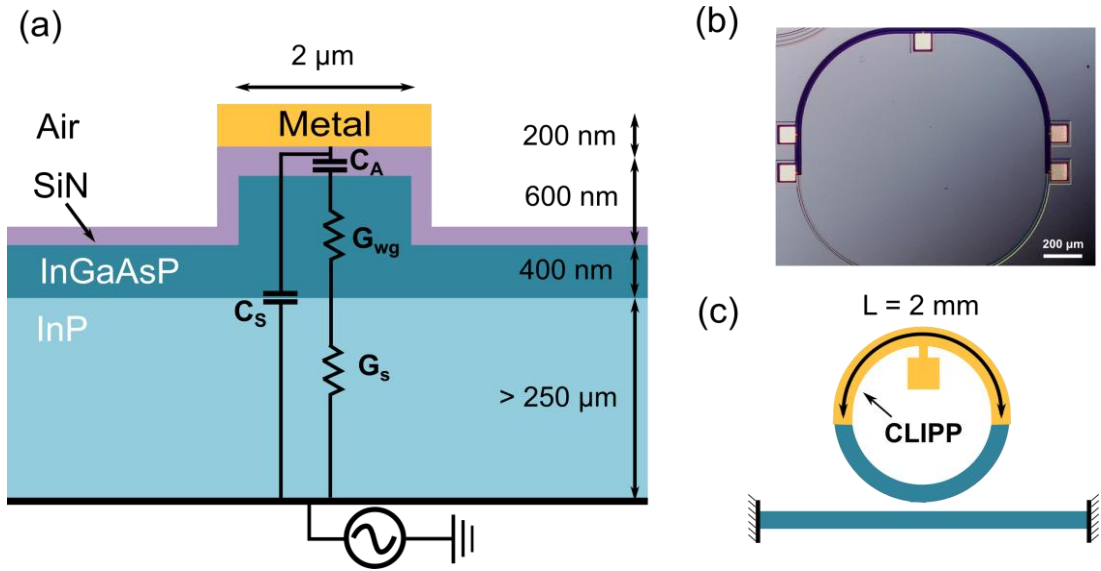

**Figure 1**: Vertical ContactLess Integrated Photonic Probe (CLIPP) on InP-based waveguide. (a) Rib waveguide cross-section with SiN isolation layer and CLIPP metal electrode. The second electrode is realized contacting the bottom of the chip. (b) Photograph and (c) schematic of the realized ring resonator with integrated vertical CLIPP.

From the electrical point of view, the small-signal lumped equivalent circuit which models the device is illustrated in figure 1(a). The SiN layer represents an insulating barrier between the top metal and the waveguide core, thus being modeled as a capacitance $C_A$, estimated for this device as about 2 pF. The conductance $G_{wg}$ of the waveguide core and $G_s$ of the substrate are in series to $C_A$. The total conductance $G = G_{wg} + G_s = \sigma L w/d$ depends on the InP conductivity σ, the length of the top electrode L, the waveguide width w and the distance between the electrodes d. G is determined by the pyramidal volume of the semiconductor in which the current flows, from the large bottom electrode to the micrometric top contact. The top section of this volume, represented by $G_{wg}$, changes according the light intensity guided by the waveguide, because of a variation of the waveguide conductivity. As already demonstrated [9], the variations of $G_{wg}$ are related to the presence of additional photo-generated carries due to surface-state absorption along the interfaces of the waveguide. For this reason, an increase of the optical power in the waveguide causes an increase of $G_{wg}$. Finally, an additional capacitor $C_S$ must be added to the equivalent model in order to take into account the direct parasitic capacitive coupling between the two electrodes. In the actual setup $C_S$ ~0.5 pF, representing the dominating term in the global admittance at electrical frequencies higher than 10 kHz. Being independent of the light intensity, the effect of $C_S$ can be removed by differential measurements.

This equivalent model is conceptually identical to that adopted for silicon waveguides [9] and the same general considerations can be applied here, even if a different dependence of the parameters to the CLIPP geometry can be observed. In the vertical CLIPP structure, once the waveguide width has been fixed (2 μm in the realized device), only the top electrode length L is available as design parameter, determining at the same time the values of $C_A$ and G. While in the coplanar scheme (SOI approach) the variation of G with light intensity scales with the inverse of the CLIPP length (i.e. with the horizontal distance between the electrodes, inversely proportional to the waveguide conductance [9]), in the vertical case the relation is opposite and ΔG ~ L. This is due to the fact that the distance between the top and bottom electrode is fixed by the waveguide cross-section and a longer CLIPP determines a higher overall conductance G. Likewise, $C_A$ increases with the same relation versus L, so that the operational frequency $f_E$ of the excitation signal used to perform the measurement becomes almost independent to the CLIPP size, significantly simplifying the design of the CLIPP.

## 3. Waveguide light monitoring

In order to probe the conductivity $G_{wg}$ of the InP waveguide, a sinusoidal excitation signal $V_E = 1$ V was applied to the large bottom electrode. The measurement requires the access capacitance $C_A$ to be negligible with respect to the waveguide impedance, that is the operational frequency $f_E$ has to be sufficiently high to shunt $C_A$, i.e. above the pole $f_E > 1/(2\pi \cdot C_A/G) \approx 1$ kHz. The top electrode, acting as sensing point, was connected to a transimpedance amplifier (TIA) with gain $10^4$ V/A and bandwidth 8 MHz. The TIA converts the current $I_E$ collected at the sensing electrode into a voltage which is fed to a lock-in demodulator (LID) for in-phase and in-quadrature measurement of the overall complex impedance Z of the structure. The measured signal is proportional, through the gain of the acquisition chain, to the admittance $Y = 1/Z$ and thus to the conductivity $G = \text{Re}\{Y\}$.

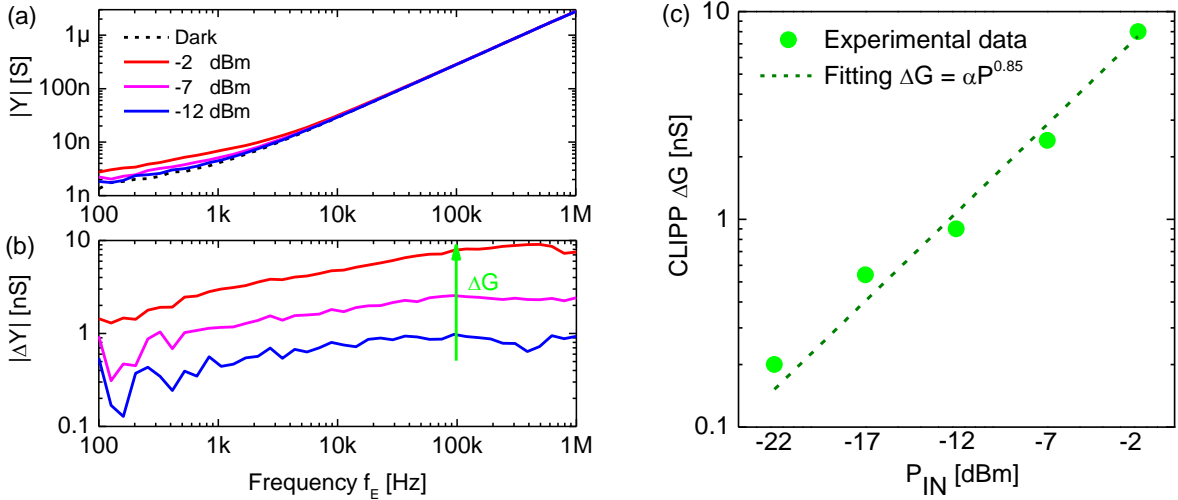

**Figure 2**: CLIPP electrical characterization. (a) Magnitude of admittance power spectra $Y(f_E)$ for different input power levels $P_{IN}$. (b) The extraction of the light-dependent increase of the conductance $\Delta G$ as difference with the dark spectrum allows to build the (c) sensor response curve.

The first characterization of the CLIPP has been performed by means of admittance spectroscopy, i.e. by measuring the total admittance Y in the range between 100 Hz and 1 MHz [figure 2(a)]. Admittance spectra were measured for a set of input power levels (from -22 dBm to -2dBm in the bus waveguide) and in the case of light absence (dark spectrum in figure 2(a), black dashed line). The latter was used as reference to obtain differential spectra $\Delta Y(f_E)$ [figure 2(b)] that allow to highlight the effect of the guided light on the waveguide admittance and to measure the variation of the conductivity $\Delta G_{wg} \approx \Delta G = \text{Re}\{\Delta Y\}$ with light intensity (assuming a constant $G_s$). A resistive region (flat admittance magnitude, minimum phase), which can be used to reliably measure $\Delta G$, can be identified in a defined frequency range: for lower frequencies the impedance of $C_A$ cannot be neglected; at higher frequencies $C_S$ dominates and amplifies any drifting difference with respect to the dark reference spectrum, thus reducing the sensitivity of the measurement. In this case, this rage is comprised between $f_E = 100$ kHz and $f_E = 1$ MHz for all the differential spectra. Figure 2(c) shows the resulting response curve of the sensor measured in this frequency region, which also presents the minimum phase for $\Delta Y$. A value of $\Delta G = 8$ nS at $P_{in} = -2$ dBm was measured and the overall response was fitted by a power law $\Delta G = \alpha P^n$ with an exponent $n = 0.85$ (dashed line).

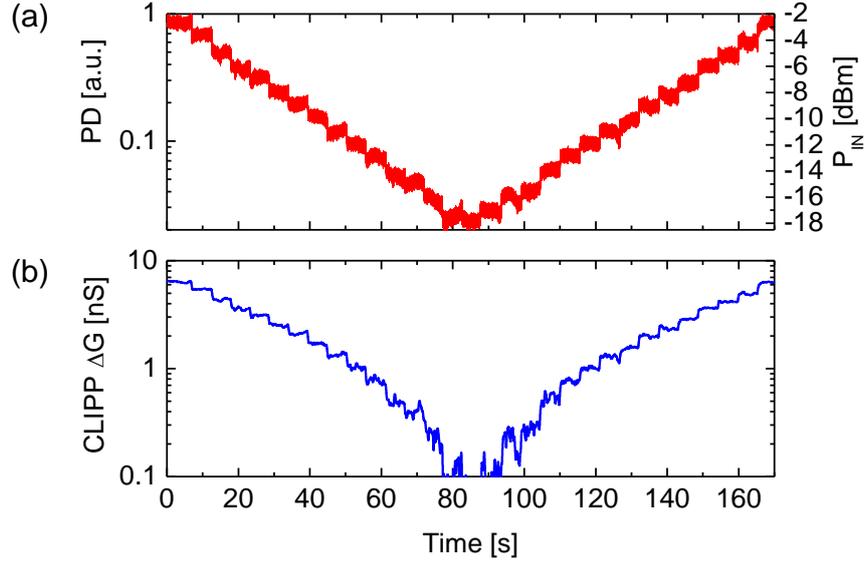

**Figure 3**: (a) Photodiode signal detected at the chip output when input power is decreased with steps of 1 dB and (b) the corresponding ΔG measured with the CLIPP.

We exploited the characterization of the sensor response reported above to test the potentiality of the realized CLIPP as in-line power monitor. To this purpose we progressively changed the optical power $P_{IN}$ in the bus waveguide from -2 dBm to -17 dBm and vice versa with discrete steps of 1 dB. Figures 3(a,b) report the corresponding signals simultaneously detected at the output photodiode and from the CLIPP. The measurement was performed with $V_E$ = 1 V, $f_E$ = 100 kHz and averaging time 1 s. As can be clearly seen, the CLIPP correctly measures all the variations applied at the input power, producing a signal consistent with the photodiode reference curve. In this conditions, the overall measured resolution of the detection system results to be ~120 pS rms and the minimum detectable power is approximately -20 dBm.

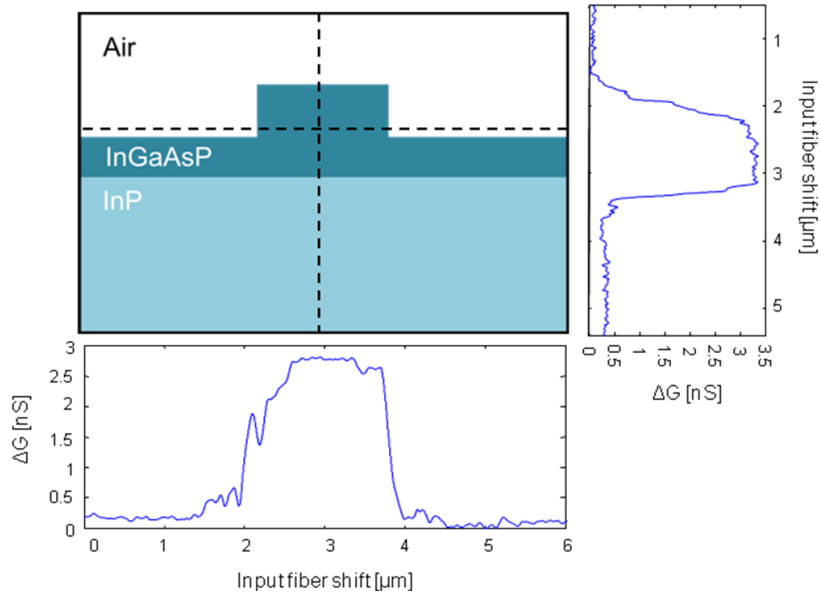

**Figure 4**: CLIPP signal versus input fiber shift in the vertical and horizontal directions. When the tapered fiber (1.7 μm waist) is not perfectly aligned with the waveguide core, the measured ΔG steeply drops.

In order to verify that the CLIPP is mainly sensitive to the light confined within the waveguide, we measured the conductance variation ΔG while moving the input tapered fiber (waist of 1.7μm) out of the perfect alignment with waveguide core. We shifted the input fiber across about 6 μm in both horizontal and vertical directions, as represented by the dashed lines in the sketch of figure 4. The crossing point represents the optimal position to minimize the insertion loss between the fiber and the bus waveguide. For both horizontal and vertical directions, a misalignment of the fiber from the optimum, and hence a reduction of the fiber-waveguide coupling efficiency, causes a steep drops of the CLIPP signal, consistent with a reduced intensity of the guided light. It is interesting to notice that in the vertical direction the alignment of the fiber in the substrate generates a ΔG slightly higher with respect to an alignment in the air on the top of the waveguide, indicating the presence of weak stray light.

### 4. Identification of shallow resonances

The realized device provides a useful test bench to demonstrate the usefulness of on-chip non-invasive monitoring of light in the analysis and control of photonic integrated circuits. Thanks to its non-invasive nature, the CLIPP can be integrated inside generic devices, such as ring resonators, enabling to access more accurate information of the device working point compared to conventional external or integrated photodetectors.

As shown in section 2, the device can be described in terms of two coupled cavities. The first one is represented by the all-pass ring resonator [blue ring in figure 5(a)]. The physical length of the ring (3.736 mm) results in a Free Spectral Range (FSR) of about 184 pm (23 GHz). The simulation of the ring transfer function performed with the commercial circuit simulator Aspic$^{TM}$ [12] showed an operational regime far from the critical coupling condition, with a finesse of the resonant cavity of 3.6 and an extinction ratio of 3 dB. The second resonant cavity of the device is the Fabry-Pérot resonator induced by the chip facets [green waveguide in figure 5(a)], with a FSR of 97 pm (12 GHz). A reflectivity of 28% was simulated for the InP waveguide termination against air since no anti reflective coating nor angled facets were used.

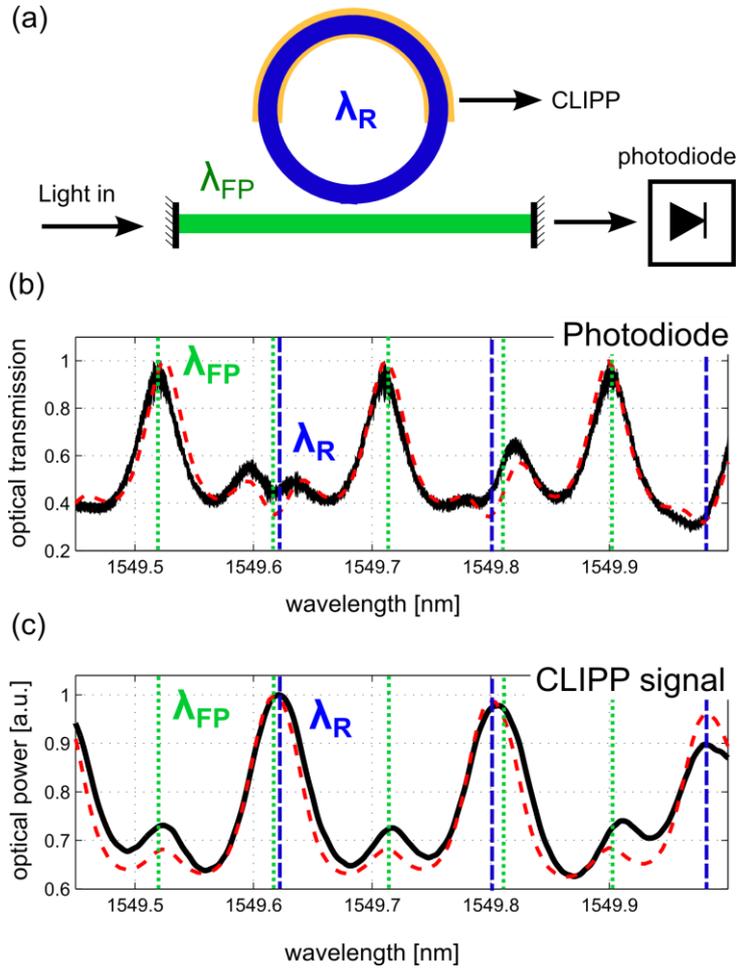

**Figure 5**: (a) Superposition of the resonances of two coupled cavities, the Fabry-Pérot ($\lambda_{FP}$) induced by the chip facets and the ring resonators ($\lambda_R$). (b) The power measured at the chip output is shown with black solid line: the fringes of the Fabry-Pérot cavity (green dotted lines) are clearly visible. On the contrary, the CLIPP (c, black line) is much more sensitive to ring resonances (blue dashed lines). The simulations (performed with Aspic[TM] [12]) of the optical power at the chip output and inside the ring (red dashed lines, (b) and (c) respectively) confirm the experimental results.

Because of the very low finesse of the ring resonator, the signal of a photodiode placed on the bus waveguide or at the output of the chip cannot be easily used to measure and possibly actively monitor the resonance frequencies of the ring (e.g. to provide a feedback to a thermal actuator). This can be clearly seen in figure 5(b). The black solid line shows the transfer function measured at the chip output with a photodiode while the red dashed line represents the corresponding, in very good agreement with the measurement. Since the FSR of the ring almost doubles that of the Fabry-Pérot, their mutual interaction causes a periodic disappearing of one the Fabry-Pérot peaks at the locations of ring notches. The expected positions $\lambda_{FP}$ for the resonance peaks of the Fabry-Pérot cavity (obtained from simulations) are marked with green dotted line. Blue dashed lines represent the (simulated) positions $\lambda_R$ of the resonant notches of the ring resonator. While the former are clearly distinguishable, the latter can be hardly identified and cannot be used to reliably measure the transfer function of the ring.

In this case is then particularly useful the possibility to detect the optical power directly inside the ring cavity, as shown in figure 5(c) with red dotted line. Although at the output port the transfer function of the ring is mostly covered by the effect of the Fabry-Pérot, the simulation of the intra-cavity optical intensity allows to easily identify the ring resonances. The measurement of the power intensity inside the ring was performed by using the CLIPP integrated in the cavity [10] through the curve of figure

2(c). The results [figure 5(c), black solid line] are almost superposed with the simulations. At the ring resonant wavelength $\lambda_R$ the effect of cavity intensity enhancement is clearly visible and allows to easily detect the transfer function of the ring resonator, unlike the measurement performed through the external photodiode.

## 5. Conclusions

We have demonstrated that the ContactLess Integrated Photonic Probe (CLIPP), originally developed on silicon waveguides and devices, can be successfully integrated also on InP-based technologies. This integration did not require any variation to the existing production process and is hence suited for the introduction in a generic platform approach.

Furthermore, the finite conductivity of the InP layer stack allowed to introduce an innovative vertical scheme for the CLIPP. This configuration requires a single sensing metal contact on top of the waveguide and exploits as a large common electrode the back side of the chip. This electrode allows all the CLIPPs to share the common excitation signal, thus reducing the area of the sensor and the space required by metal pads and electrical connections.

We reported the functionality of the CLIPP as a power monitor able to detect light intensity within the waveguide. The integration inside a ring resonator has been used as a severe test condition to monitor the spectral response of the ring even if highly distorted by an external Fabry-Pérot cavity.

## 6. Acknowledgments

The authors gratefully acknowledge Franscisco M. Soares, Moritz Baier and Norbert Grote of the Fraunhofer Heinrich Hertz Institut (Berlin) for the fabrication of the devices. This work was partially supported by the European Community's Seventh Framework Programme FP7/2007-2013 under Grant ICT 257210 (PARADIGM) and by the European project BBOI.